\documentclass[12pt,a4paper]{article}

\usepackage{amssymb}
\usepackage{setspace}
\usepackage{indentfirst}

\usepackage{graphicx}

\usepackage{url}

\begin{document}

\begin{titlepage}
\begin{center}

\vspace*{25mm}

\begin{spacing}{1.7}
{\LARGE\bf Polarization of CMB and possible time dependence of dark energy}
\end{spacing}

\vspace*{25mm}

{\large
Noriaki Kitazawa
}
\vspace{10mm}

Department of Physics, Tokyo Metropolitan University,\\
Hachioji, Tokyo 192-0397, Japan\\
e-mail: noriaki.kitazawa@tmu.ac.jp

\vspace*{25mm}

\begin{abstract}
Dark energy has been introduced to explain
 the present accelerating expansion of the universe.
In the $\Lambda$CDM model, the present standard model of cosmology,
 dark energy is described as a cosmological constant
 which is time independent.
The possibility of the time dependence of dark energy
 has been investigated to obtain deeper understanding of it
 by looking redshift-magnitude relation of type Ia supernova, for example.
We investigate the constraints to the time dependence of dark energy
 with several phenomenological models
 by fitting their parameters to the redshift-magnitude relation
 of the Pantheon supernova catalog,
 and confirm that little time dependence can not be excluded
 by the goodness-of-fit criterion only.
Such a non-trivial time dependence
 causes different expansion rate in the period of reionization,
 which affects the low-$\ell$ polarization power spectrum of CMB. 
We investigate the possibility
 to detect the effect in future probe like LiteBIRD, for example.
We find that
 the low-$\ell$ EE polarization power spectrum is enhanced in general,
 but it will be difficult to be detected
 as far as looking the EE polarization power spectrum only
 because of the limitation by cosmic variance.
\end{abstract}

\end{center}
\end{titlepage}

\doublespacing

\section{Introduction}
\label{sec:introduction}

The nature of dark energy,
 which realizes the present accelerating expansion of the universe
 \cite{Riess:1998cb,Perlmutter:1998np}, is unclear.
Though it can be described as a cosmological constant
 in the $\Lambda$CDM model, the present standard model of cosmology,
 it is simply a possible parameter and the origin is unknown.
The cosmological constant can cause accelerating expansion of the universe
 with time-independent expansion rate.
If dark energy is some kind of vacuum energy or some effect of modified gravity
 (for reviews, see \cite{Silvestri:2009hh,Clifton:2011jh,Joyce:2014kja},
  and also see \cite{Raveri:2019mxg} for a recent effort),
 the expansion rate can be time-dependent.

Since an absolutely convincing and well-motivated origin of dark energy has not been found yet,
 some phenomenological models of dark energy beyond the cosmological constant
 have been introduced, and they have been examined by cosmological observations.
A typical phenomenological model is Chevallier-Polarski-Linder (CPL) model
 \cite{Chevallier:2000qy,Linder:2002et}
 of non-trivial time dependence in the equation of state of dark energy $p = w \rho$ as
\begin{equation}
 w = w_0 + w_a (1-a(t)),
\end{equation}
 where $w_0$ and $w_a$ are numerical constants, namely parameters,
 and $a(t)$ is the scale factor
 assuming flat Friedmann-Lema\^{i}tre-Robertson-Walker metric
 with the normalization of present value of the scale factor $a(t_0)=1$.
The value of $w$ is $w_0$ at present,
 and it changes linearly to $w_0+w_a$ at very early time. 
The set of values of $w_0=-1$ and $w_a=0$ corresponds to the cosmological constant,
 or the $\Lambda$CDM model.
The model with $w_a=0$ keeping $w=w_0$ as a free parameter is called $w$CDM model.
The typical constraints to these parameters are
\begin{equation}
 w_0 = -0.939 \pm 0.073,
\qquad
 w_a = -0.31^{+ 0.28}_{-0.24}
\end{equation}
 for CPL model, and
\begin{equation}
 w = -1.020 \pm 0.027
\end{equation}
 for $w$CDM model
 by using PLANCK CMB, Pantheon supernova, SDSS BAO+RSD and DES 3$\times$2pt data
 \cite{Alam:2020sor}.
Though these results indicates that
 the $\Lambda$CDM model is consistent with present data,
 the time dependence of $w$, namely $w_a$ has not been strongly constrained yet.
It seems to be required
 to find some other way to constrain further the time dependence of $w$.

The non-trivial time dependence of dark energy
 changes the way of expansion of the universe
 not only in the era of dark energy dominance ($z < 0.3$),
 but also in the era of reionization ($6 < z < 10$),
 though the dark energy contribution is not dominant in the latter era.
Furthermore,
 the change of the values of cosmological parameters, like matter density $\Omega_m$,
 through the fit with data by observations in the region of $z < 0.3$,
 where the effect of the non-trivial time dependence of dark energy is large,
 also changes the way of expansion in the era of reionization.
This change affects
 the large-scale (low-$\ell$) power spectrum of the polarization of CMB,
 since the polarization is produced by the Thomson scattering with free electrons
 which are produced in the reionization process
 (see \cite{Kosowsky:1994cy,Hu:1997hv,Cabella:2004mk} for reviews).
In this article we examine the possibility
 whether future precise measurements of the large-scale polarization power spectrum
 can give some constraints to the time dependence of dark energy.

We introduce the following two phenomenological models further,
 since the extrapolation of the CPL model to larger redshift region,
 beyond the era of dark energy dominance, is non-trivial and may be dangerous.
The model of simple Taylor expansion
\begin{equation}
 w = w_0 + w_1 \, a(t) + \frac{1}{2} w_2 \, a(t)^2
\end{equation}
 with $w_0=-1$ naturally arrives at the case of the cosmological constant in early time,
 where $w_1$ and $w_2$ are free parameters.
We call this model Taylor expansion model.
We consider another simpler model with $w_2=0$ keeping $w_1$ as a free parameter,
 which we call linearCDM model.
This model can be understood as a constrained CPL model.

In the next section
 we are going to obtain best-fit values of the parameters in these four models
 under the observed redshift-magnitude relation of type Ia supernova.
We use the Pantheon catalog of type Ia supernova \cite{Scolnic:2017caz}
 \footnote{The data are open in public at \url{http://dx.doi.org/10.17909/T95Q4X}.}.
In the analysis the binned data (40 redshift bins) are used.
We use vary simple likelihood functions,
 since our aim is not to set the precise constraints to these parameters,
 but to obtain typical reference values of parameters
 for the analyses of CMB polarization power spectrum.
In section \ref{sec:polarization}
 we calculate low-$\ell$ EE polarization power spectrum in each phenomenological model
 and compare it to that in $\Lambda$CDM model.
The power spectrum is numerically calculated by using CAMB \cite{Lewis:1999bs} code.
In the last section we conclude.

\section{Model parameters and redshift-magnitude relation}
\label{sec:parameters}

The distance modulus of an object is defined by $\mu = m - M$,
 where $m$ is the apparent magnitude of the object
 and $M$ is the absolute magnitude assuming that the object were at distance $10$pc
 from the observer.
The relation between the luminosity distance $d_L$ to the object and the distance modulus
 is $\mu = 5 \log (d_L [{\rm Mpc}]) + 25$,
 where $d_L$ is measured in Mpc and the base of the logarithm is $10$.
Consider the case that the light from the object is observed with cosmological redshift $z$.
The luminosity distance to the object is given by
\begin{equation}
 d_L(z) = \frac{1+z}{H_0} \int_0^z dz' \frac{1}{E(z')},
\end{equation}
 where $H_0$ is the Hubble constant and $E(z) \equiv H(z)/H_0$
 with $H(z)$ is the Hubble parameter as a function of redshift.
The function $E(z)$ is described by model parameters:
\begin{equation}
 E(z) = \sqrt{\Omega_m (1+z)^3 + (1-\Omega_m)(1+z)^{3(1+w_0+w_a)}e^{-3w_a\frac{z}{1+z}}}
\label{E-CPL}
\end{equation}
 for CPL and $w$CDM models, and
\begin{equation}
 E(z) = \sqrt{\Omega_m (1+z)^3
  + (1-\Omega_m)e^{3w_1\left(1-\frac{1}{1+z}\right)+\frac{3}{4}w_2\left(1-\frac{1}{(1+z)^2}\right)}}
\label{E-Taylor}
\end{equation}
 for Taylor expansion and linearCDM models.
Here, we simply neglect the contribution of radiation,
 which is very small in the era of redshifts that we will investigate,
 and set the energy density of dark energy $\Omega_{\rm DE}=1-\Omega_m$
 assuming flat universe.
Then, the redshift-magnitude relation of type Ia supernova is described by
\begin{equation}
 m(z) = 5 \log \left( (1+z) \int_0^z dz' \frac{1}{E(z')} \right) - 5 \, a_{\rm SN},
\label{redshift-magnitude-relation}
\end{equation}
 where $a_{\rm SN} \equiv \log(H_0 [{\rm Mpc}^{-1}]) - 5 - M/5$
 is a constant for ideal type Ia supernova with a unique absolute magnitude.
In fact
 the determination of Hubble constant from type Ia supernova data is achieved in principle
 by extracting $a_{\rm SN}$ from data by fitting redshift-magnitude relation
 with the knowledge of a unique absolute magnitudes of type Ia supernova
 (see \cite{Efstathiou:2021ocp} for a short review).
In this article we do not determine the Hubble constant,
 but extracting model parameters as well as $a_{\rm SN}$ by fitting redshift-magnitude relation
 with eq.(\ref{redshift-magnitude-relation}).

\begin{figure}[t]
\centering
\includegraphics[width=45mm]{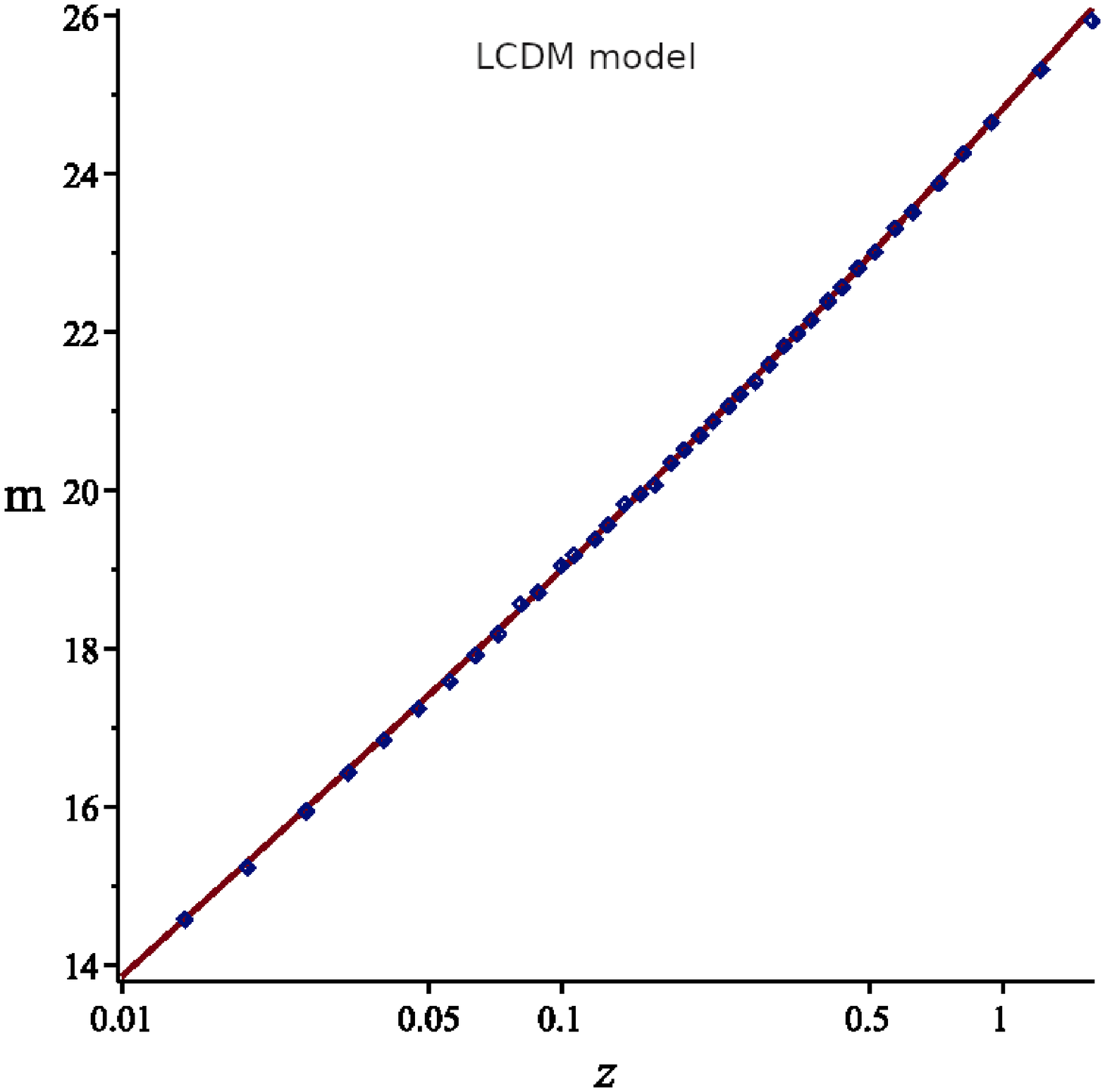}
\quad
\includegraphics[width=45mm]{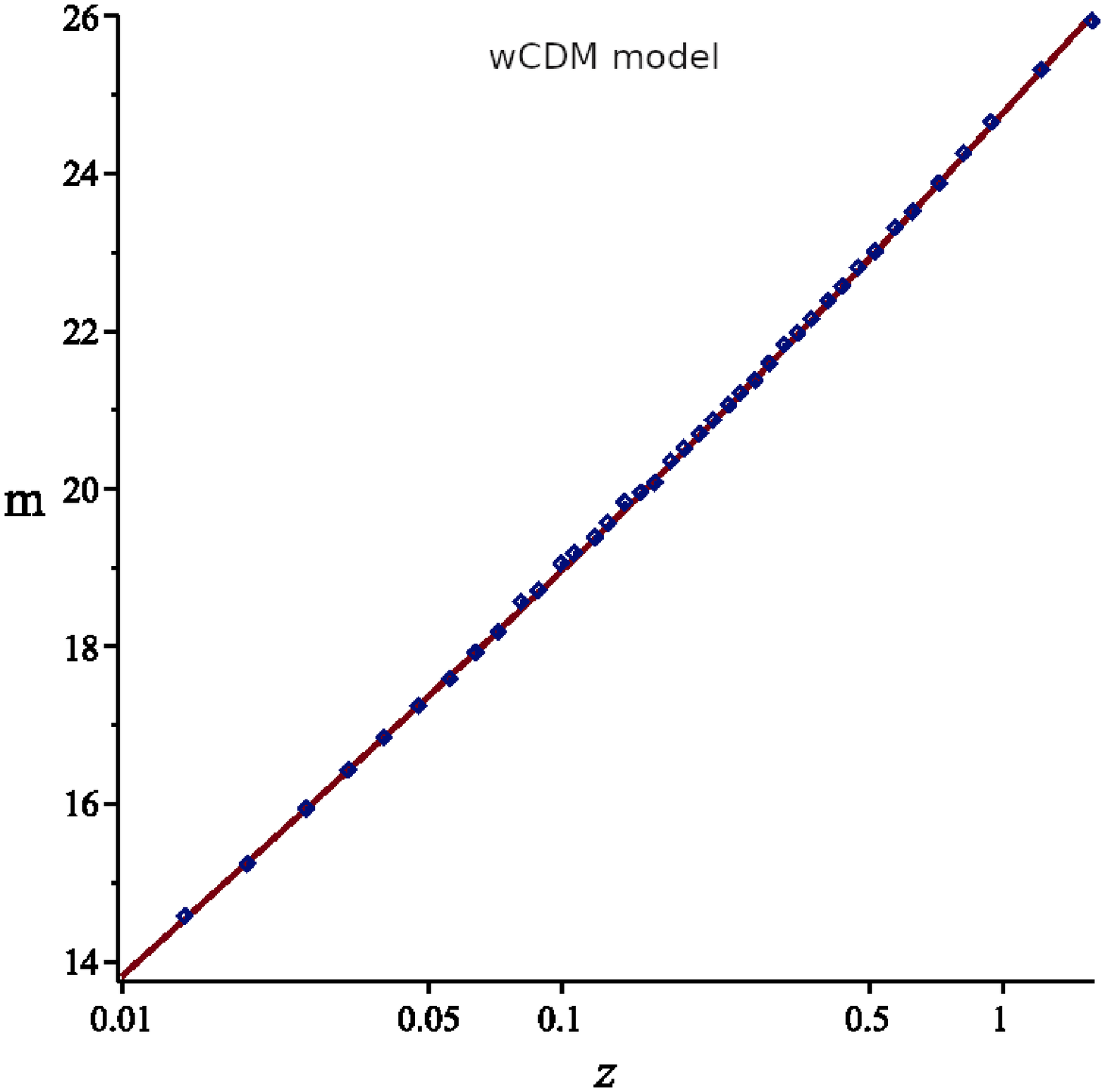}
\quad
\includegraphics[width=45mm]{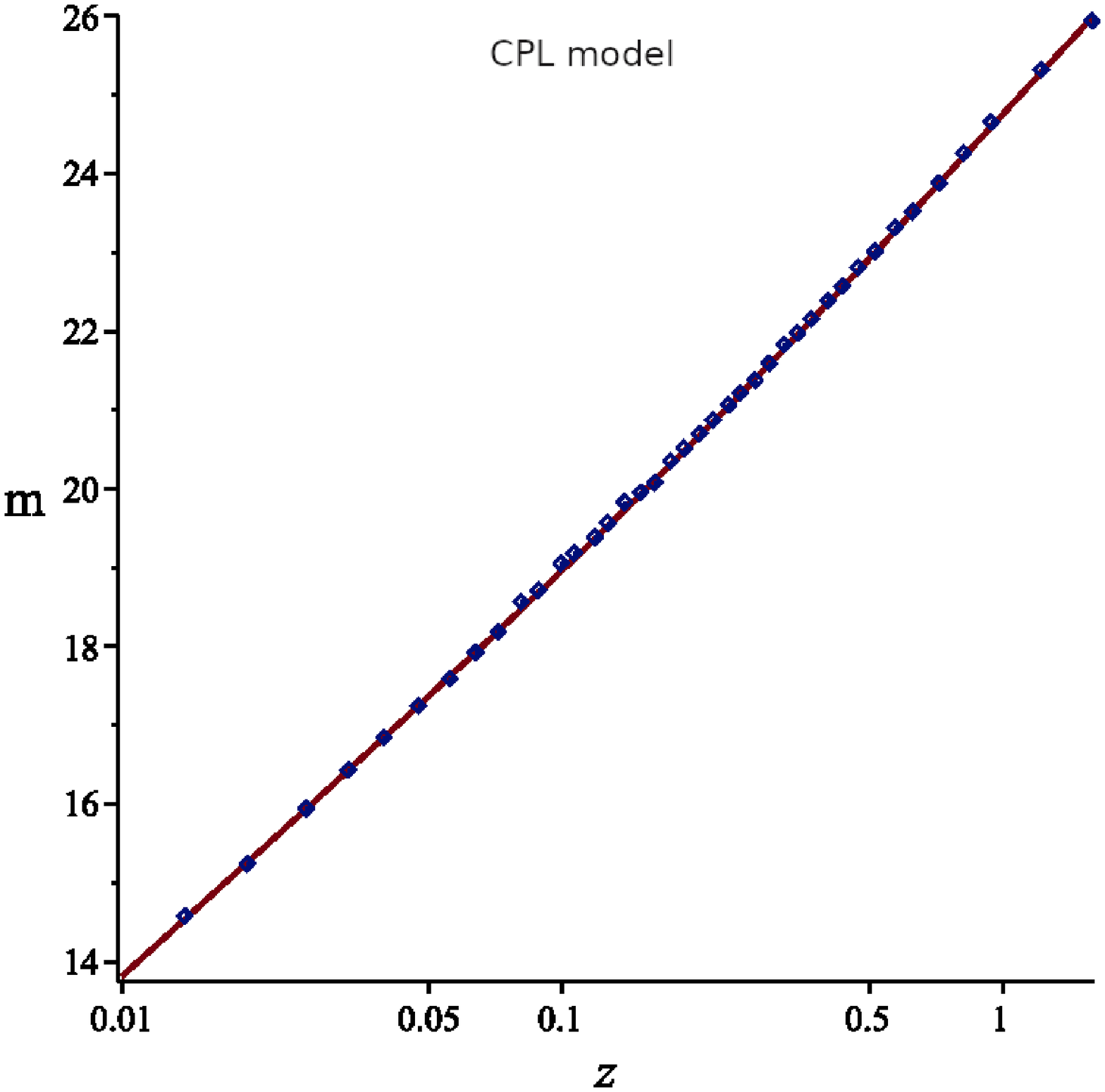}
\caption{
The Pantheon redshift-magnitude relation (40 redshift bins) fitted by three models:
 $\Lambda$CDM, $w$CDM and CPL models from left to right.
The Pantheon data are displayed as dots, which include error bars,
 and the lines are best fit results of corresponding three models.
All the figures look the same and can not be distinguished with eyes.
}
\label{fig:fit-redshift-magnitide}
\end{figure}
\begin{table}[t]
\centering
\caption{
The best fit model parameters with Pantheon binned data.
In addition to the best fit values of parameters in each model,
 minimum values of $\chi^2$ ($\chi^2_{\rm min}$)
 and corresponding reduced $\chi^2$ ($\chi^2_{\rm min}/\nu$ with degrees of freedom $\nu$) are also given.
The numbers in brackets mean fixed values.
}
\label{table:fit-binned-data}
\begin{tabular}{lccccccc}
\hline
 Model & $\Omega_m$ & $w_0$ & $w_a$ & $w_1$ & $w_2$ & $\chi^2_{\rm min}$ & $\chi^2_{\rm min}/\nu$ \\
\hline\hline
 $\Lambda$CDM & 0.28 & (-1) & - & - & - & 49.9 & 1.28 \\
 $w$CDM       & 0.32 & -1.1 & - & - & - & 47.6 & 1.25 \\
 CPL          & 0.38 & -1.2 & -0.80 & - & - & 47.5 & 1.28 \\
 linearCDM    & 0.32 & (-1) & - & -0.11 & - & 47.6 & 1.25 \\
 Taylor       & 0.35 & (-1) & - & -0.7 & 1.1 & 47.3 & 1.28 \\
\hline
\end{tabular}
\end{table}
The results of fitting the Pantheon binned data of redshift-magnitude relation
 by three models, $\Lambda$CDM, $w$CDM and CPL models, are shown in Fig.~\ref{fig:fit-redshift-magnitide}.
The $\chi^2$ statistics for fitting is defined as
\begin{equation}
 \chi^2 \equiv \sum_{i=1}^{40} \frac{(m_i - m(z_i))^2}{(\sigma_i)^2},
\end{equation}
 where $m_i$, $z_i$ and $\sigma_i$
 are Pantheon data corresponding to the magnitudes, redshifts and errors of the magnitude,
 respectively.
We adopt this simple one,
 since our aim is not to set precise constraints to model parameters,
 but to obtain typical reference values of model parameters
 for the analyses of CMB polarization power spectrum.
Though each line in Fig.~\ref{fig:fit-redshift-magnitide}
 is corresponding to the best fit values of model parameters,
 which make the value of $\chi^2$ minimum,
 it is impossible to distinguish each of them with eyes,
 since errors of the data are very small.
The same happens for other two models, linearCDM and Taylor expansion models.

The summary of best fit values of parameters of each model
 is given in Table~\ref{table:fit-binned-data}.
The matter energy density parameter $\Omega_m$
 is considered as a parameter in $\Lambda$CDM model and also in other models.
The ranges of scan of parameters are:
 $0.10 \leqq \Omega_m \leqq 0.50$, $-1.5 \leqq w_0 \leqq -0.5$,
 $-1.0 \leqq w_a \leqq 1.0$, $-1.2 \leqq w_1 \leqq 1.2$, $-1.2 \leqq w_2 \leqq 1.2$
 and $-5.0 \leqq a_{\rm SN} \leqq -3.0$.
The best fit value of $a_{\rm SN}$ is equally $-4.76$ for all the models,
 that value correspond to $H_0 = 73.2 \, {\rm [km/s \, Mpc]}$ with $M=-19.263 \, {\rm[mag]}$.
The best fit values of $\Omega_m$
 in the models with one parameter, $w$CDM and linearCDM models,
 are larger than that in $\Lambda$CDM model.
The same happens in the models with two parameters, CPL and Taylor expansion models,
 in which the best fit values of $\Omega_m$ are larger than those in one parameter models.
In other words,
 better fitting the curve of redshift-magnitude relation by many parameters
 results larger value of $\Omega_m$.
However, as expected in Fig.~\ref{fig:fit-redshift-magnitide},
 goodness-of-fit (reduced $\chi^2$) for each model is almost the same,
 and we can not choose the best model.
It may indicate some limitation of this method to constrain model parameters,
 and we may need some new observables to go further.
We are going to investigate the possibility of CMB polarization power spectrum in the next section.

\section{Polarization of CMB and time-dependent dark energy}
\label{sec:polarization}

The time dependence of dark energy
 modifies the way of expansion of the universe in the period of reionization ($6 < z < 10$),
 and it affects the CMB polarization at low-$\ell$ (large scales),
 since the polarization is produced through the scattering with electrons
 which are produced in the process of reionization.
Here, we concentrate on the EE polarization power spectrum at low-$\ell$,
 on which the cosmic-variance-limited measurement is expected in near future.
Such a precise measurement
 may give further constraint or discovery of the time dependence of dark energy.

The EE polarization power spectrum
 $D^{EE}_\ell = \ell (\ell+1) C^{EE}_\ell/2\pi$ at low-$\ell$ for each model
 is numerically calculated using CAMB code with a set of basic cosmological parameters
 obtained by PLANCK \cite{Aghanim:2018eyx} :
 $\Omega_b h^2 = 0.022$, $\tau = 0.054$, $A_s=2\times10^{-9}$ and $n_s=0.965$.
On $\Omega_c h^2$ we use the values of $\Omega_m$ by our fits
 with Hubble parameter $H_0=73.2 \, {\rm [km/s \, Mpc]}$,
 which is obtained from a supernova observation \cite{Riess:2020fzl}.
We can include a non-trivial equation of state of dark energy $w(a)$ in CAMB code
 using the option which sets $w(a)$ from numerical values.

\begin{figure}[t]
\centering
\includegraphics[width=40mm]{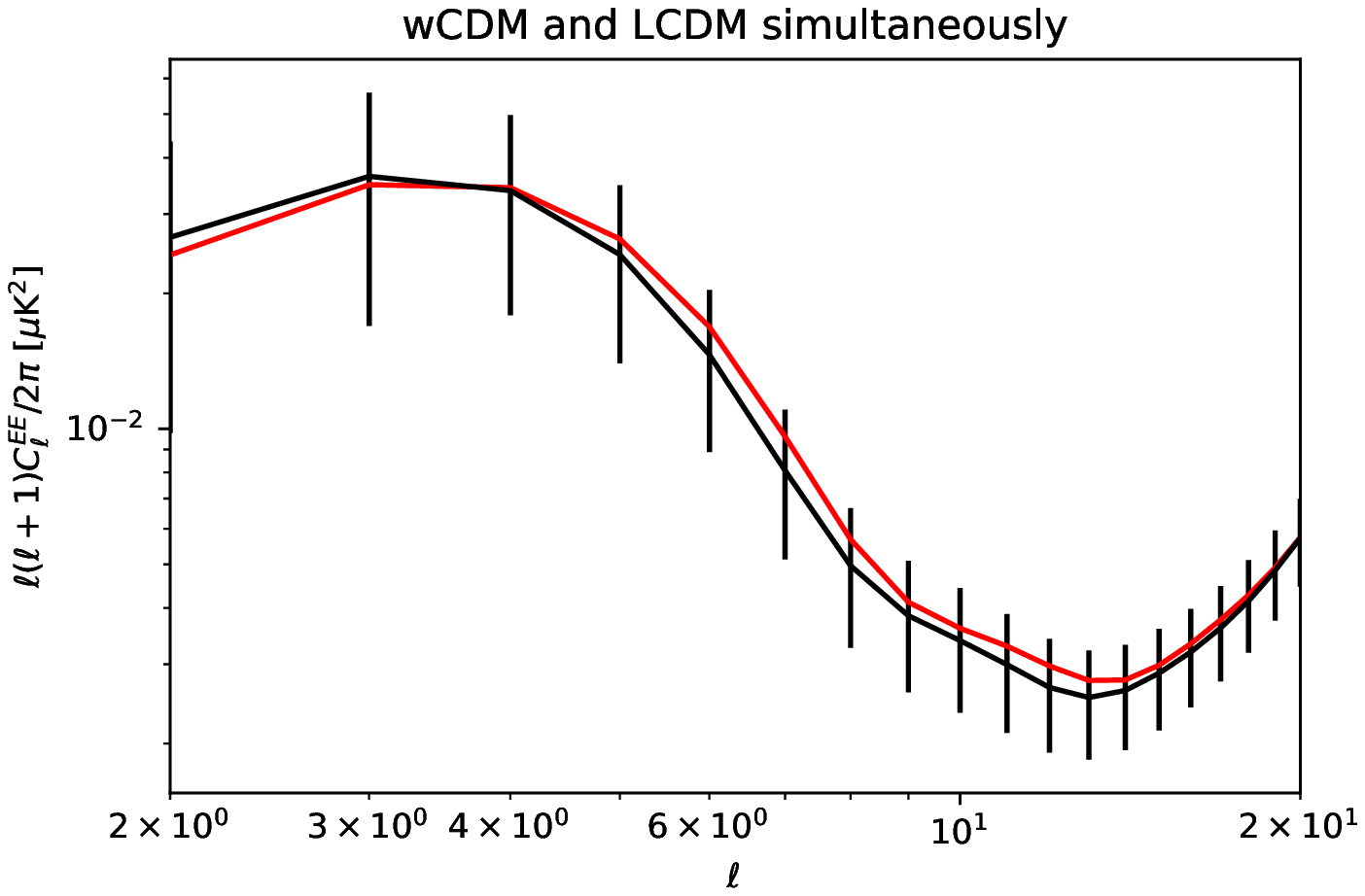}
\includegraphics[width=40mm]{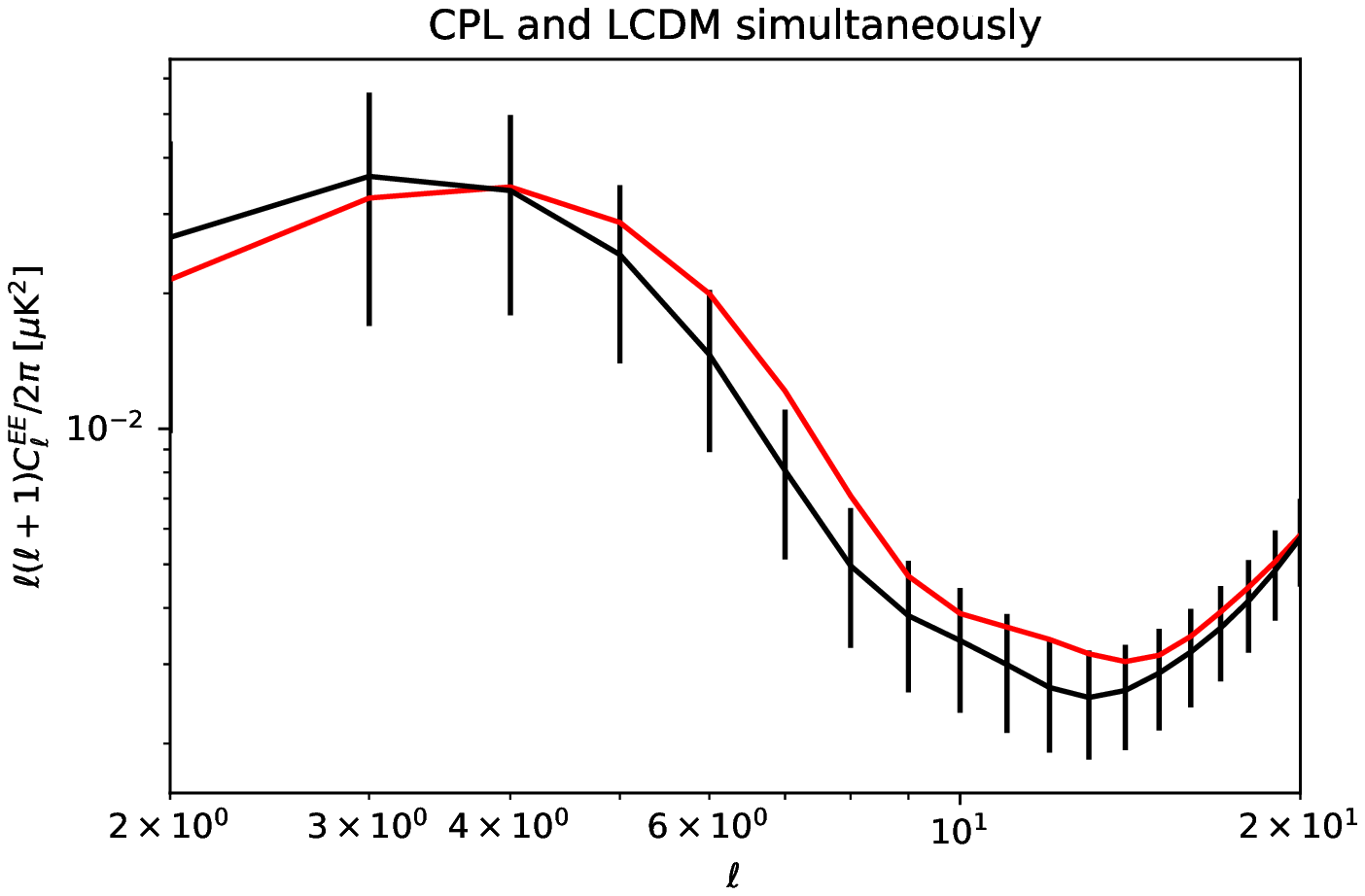}
\includegraphics[width=40mm]{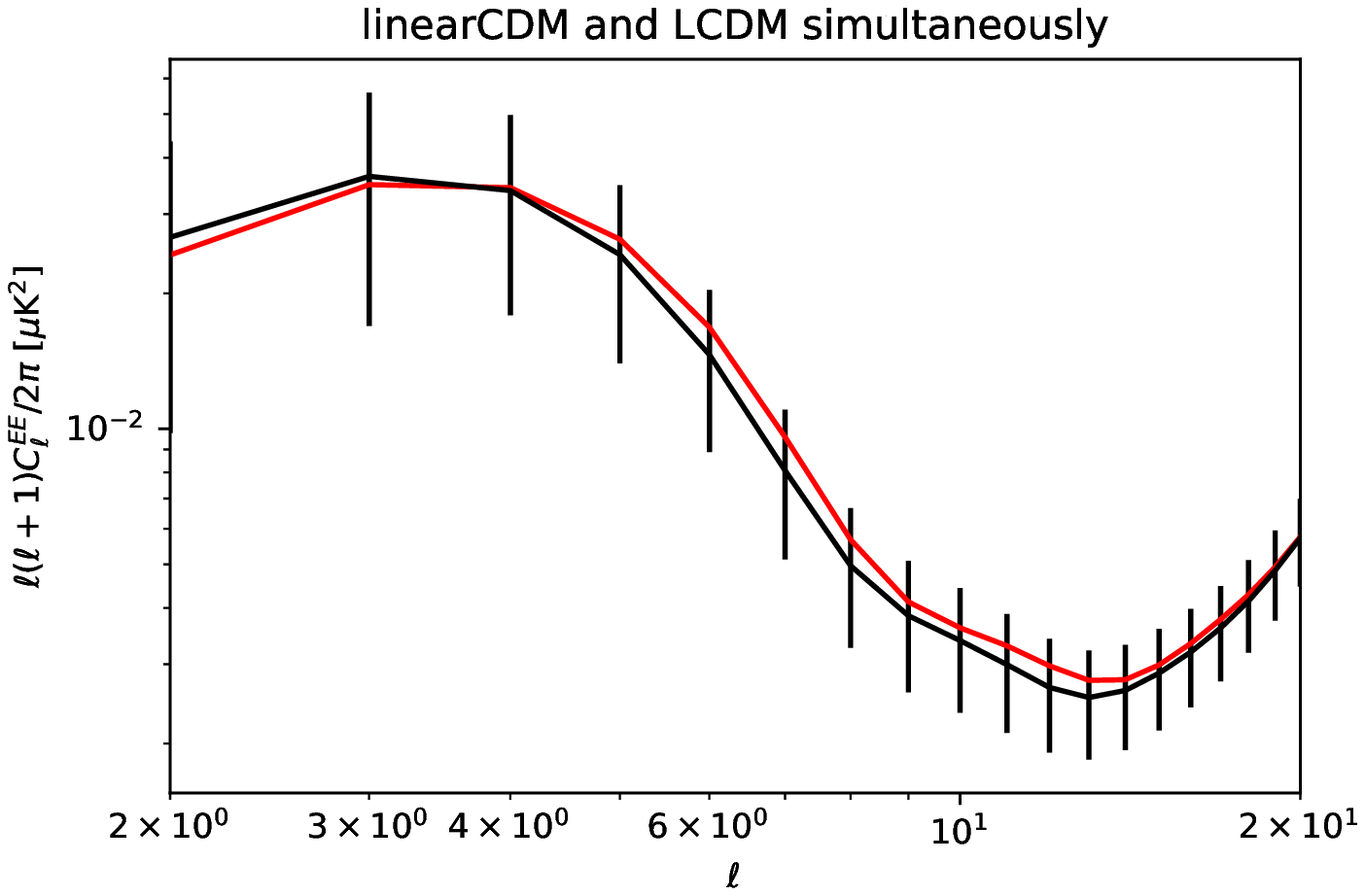}
\includegraphics[width=40mm]{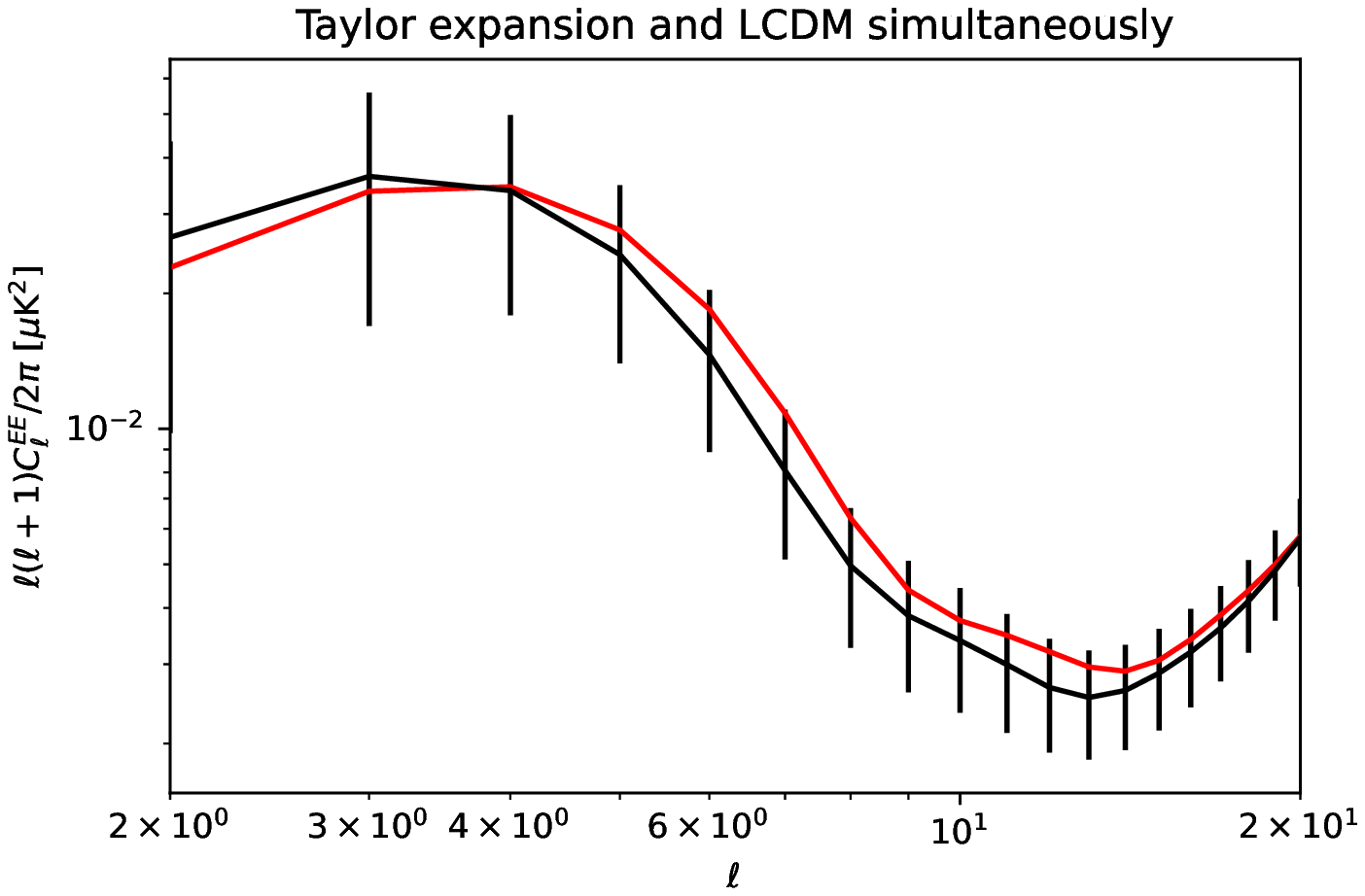}
\caption{
EE polarization power spectrum $D^{EE}_\ell = \ell (\ell+1) C^{EE}_\ell/2\pi$
 with Pantheon binned data
 for $w$CDM, CPL, linearCDM and Taylor expansion models, from left to right, respectively.
Each figure includes two lines: predictions of $\Lambda$CDM model (in black)
 and of phenomenological model (in red).
The vertical lines in each figures
 show cosmic variance limited errors in the $\Lambda$CDM model.
All the phenomenological models predict larger power for larger $\ell$.
}
\label{fig:Pantheon-D_ell}
\end{figure}
\begin{figure}[t]
\centering
\includegraphics[width=40mm]{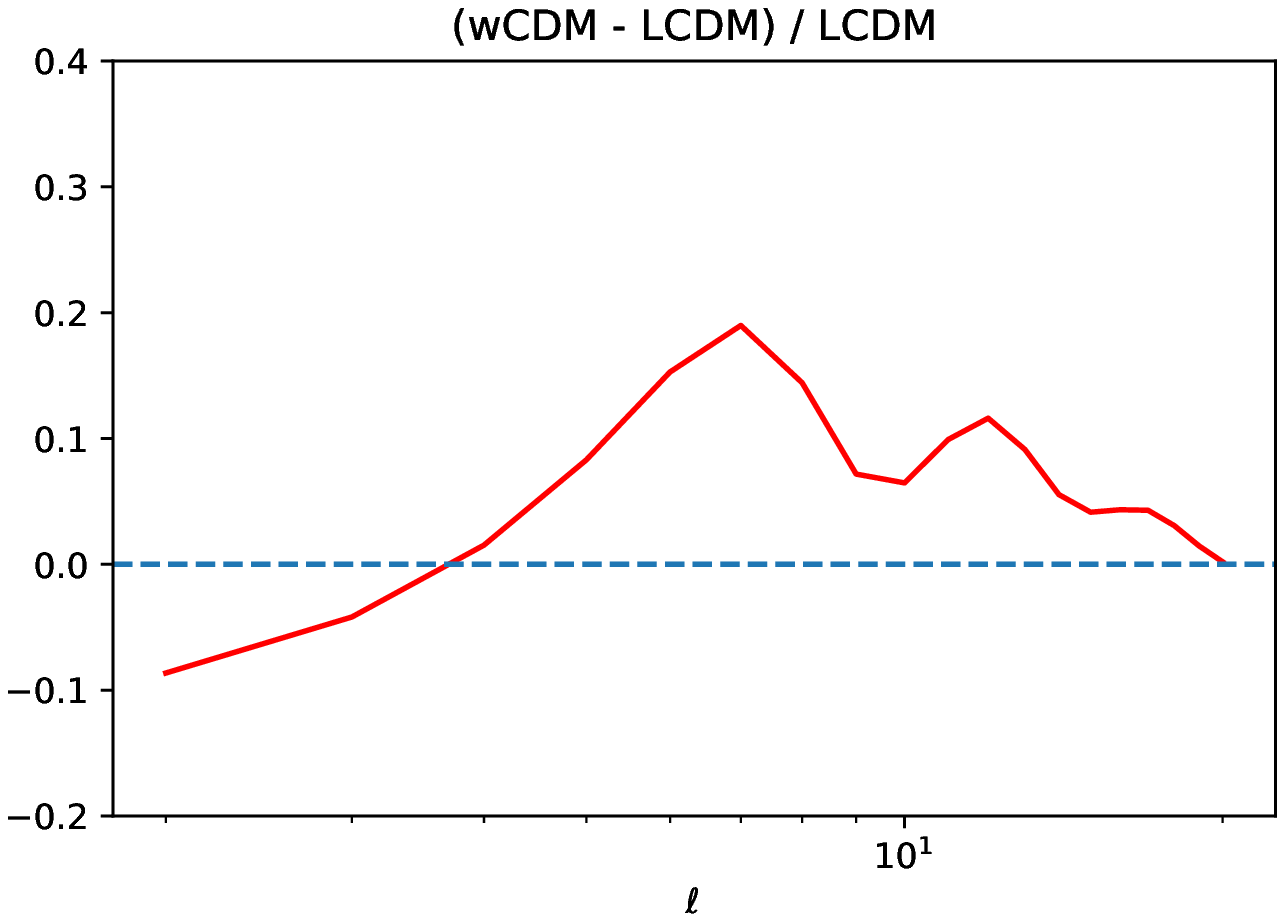}
\includegraphics[width=40mm]{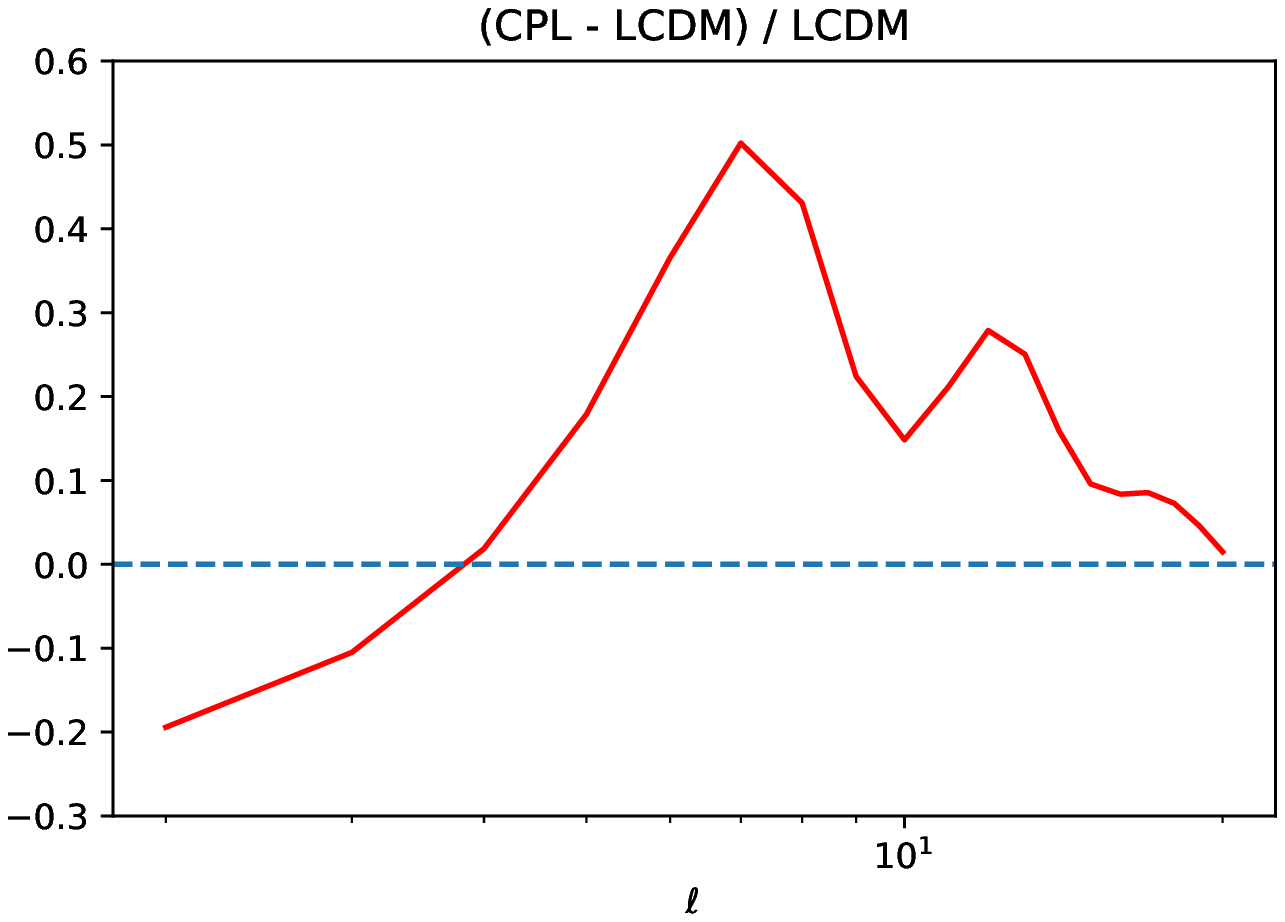}
\includegraphics[width=40mm]{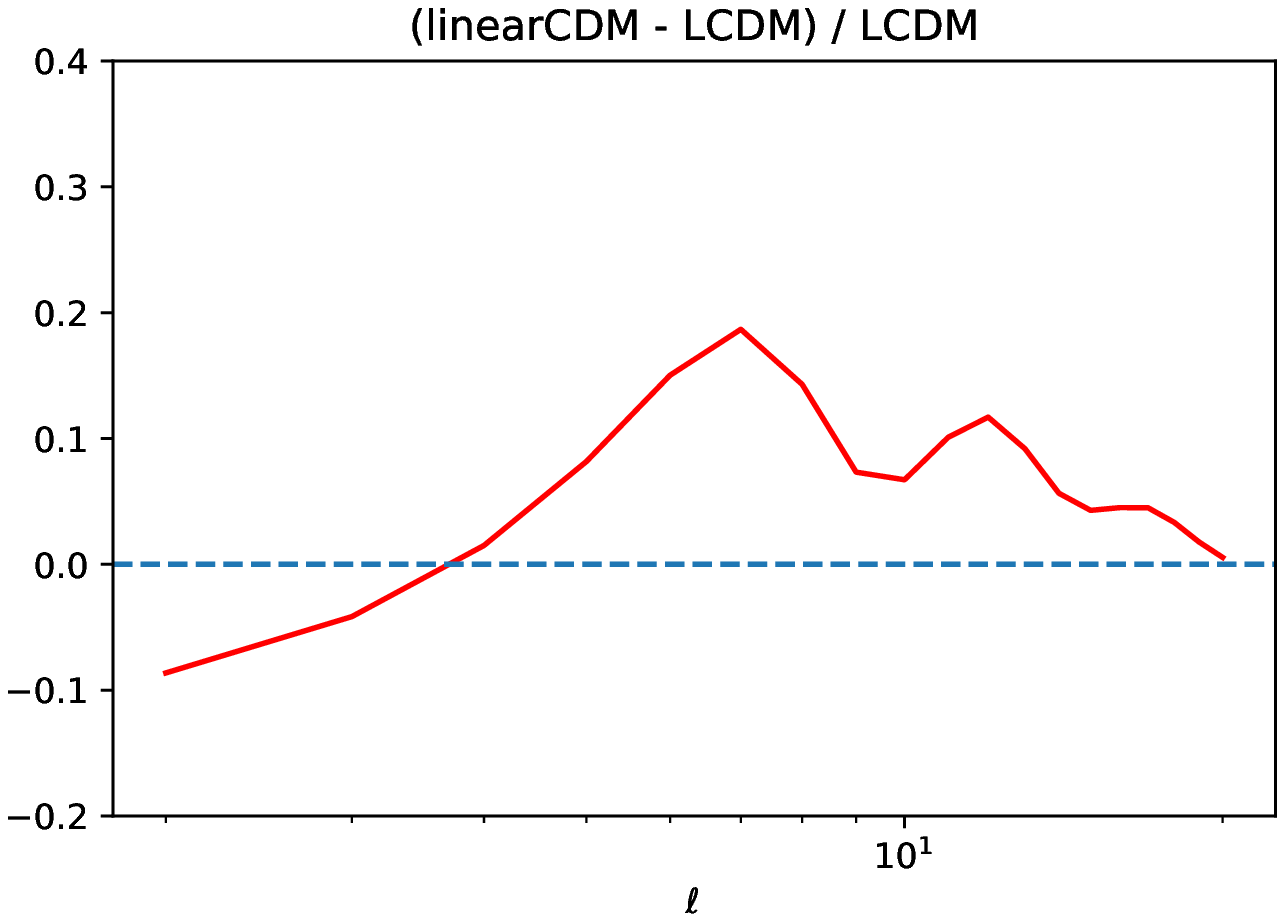}
\includegraphics[width=40mm]{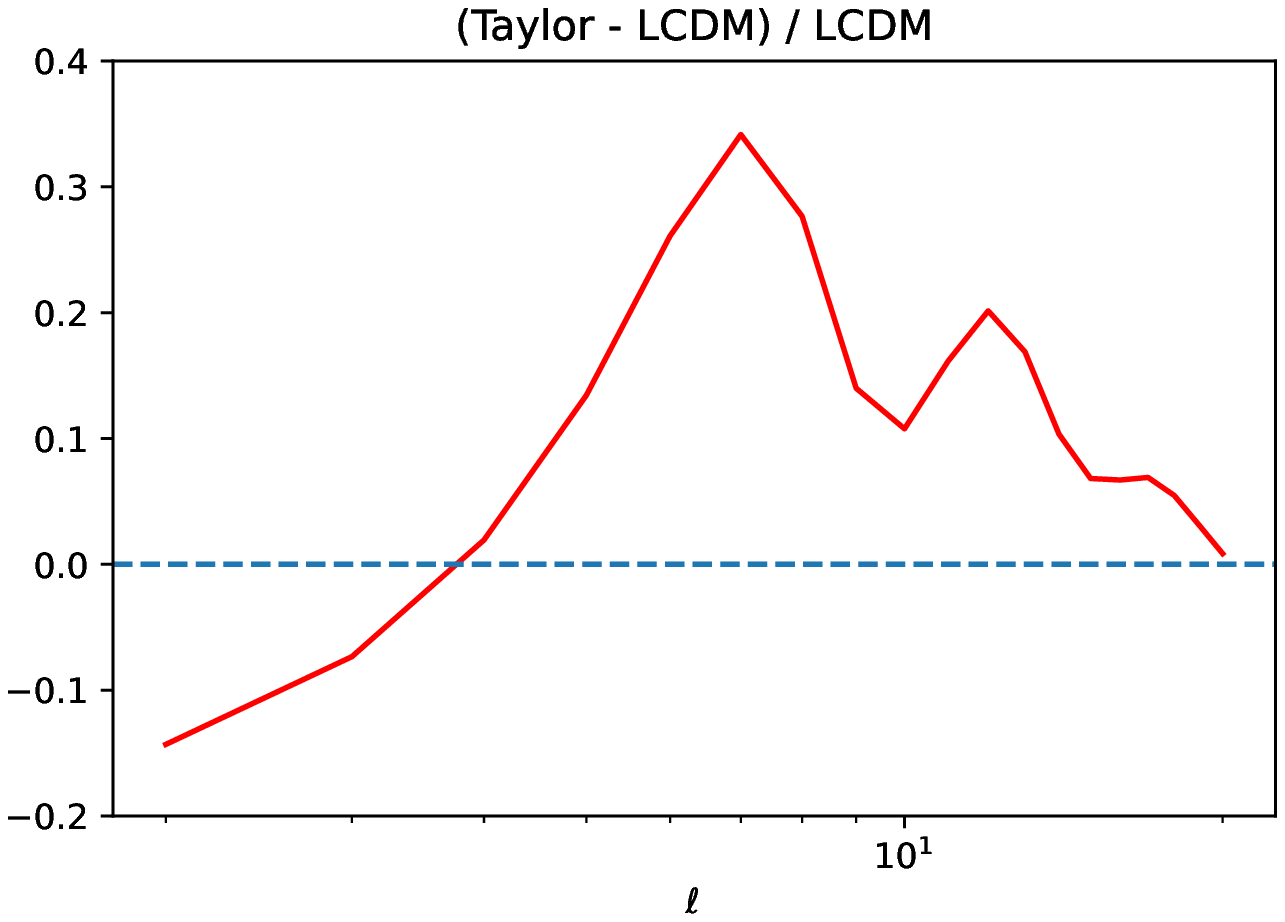}
\caption{
The differences of EE polarization power spectra
 between in phenomenological models and $\Lambda$CDM model:
 $(D^{EE, \rm model}_\ell - D^{EE, \Lambda{\rm CDM}}_\ell)/D^{EE, \Lambda{\rm CDM}}_\ell$
 with the Pantheon binned data
 for $w$CDM, CPL, linearCDM and Taylor expansion models, from left to right, respectively.
There is a common qualitative pattern of $\ell$-dependent deviation
 from the prediction of $\Lambda$CDM model.
}
\label{fig:Pantheon-diff}
\end{figure}

The results with the best fit model parameters
 to the Pantheon binned data (Table~\ref{table:fit-binned-data})
 are shown in Figs.~\ref{fig:Pantheon-D_ell} and \ref{fig:Pantheon-diff}.
The black and red lines in each panel in Fig.~\ref{fig:Pantheon-D_ell} 
 are EE polarization power spectra ($2 < \ell < 20$)
 in the $\Lambda$CDM model and phenomenological model, respectively.
The difference of EE polarization power spectrum in each model from that in the $\Lambda$CDM model
 is shown in Fig.~\ref{fig:Pantheon-diff}.
We see that
 the power are enhanced about order of 10\%
 in all the phenomenological models in the range of $4 < \ell < 20$.
This effect has been pointed out in \cite{Kitazawa:2020qdx}
 by semi-analytic calculations with some approximations.
Fig.~\ref{fig:Pantheon-diff} indicates that
 the power spectra in $w$CDM and linearCDM model are almost the same.
Note that the scale of vertical axes are different
 in the plots for CPL and Taylor expansion models in Fig.~\ref{fig:Pantheon-diff},
 though the shapes are almost the same.
The largest enhancement of 50\% happens in CPL model at $\ell=7$.

\begin{figure}[t]
\centering
\includegraphics[width=40mm]{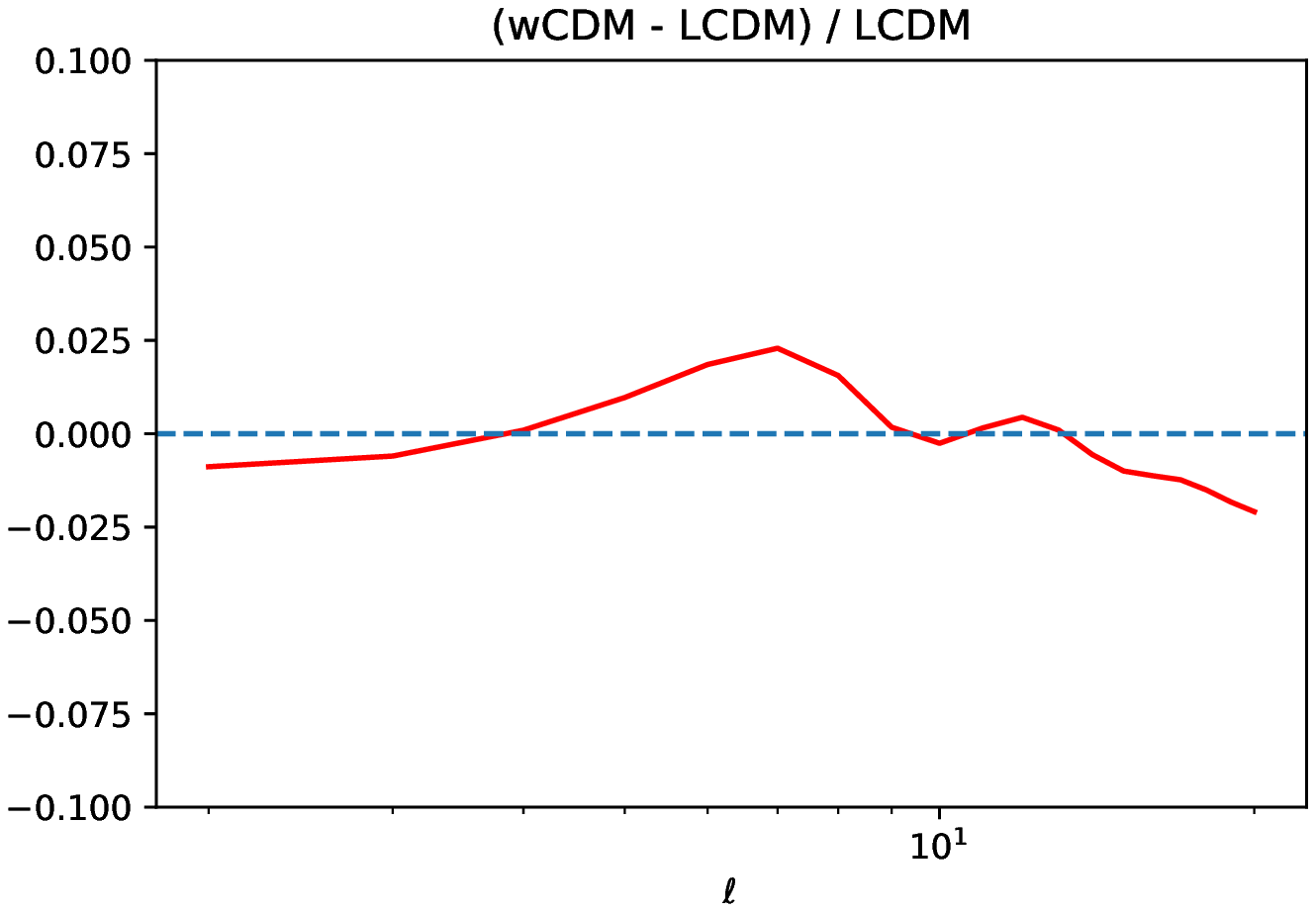}
\includegraphics[width=40mm]{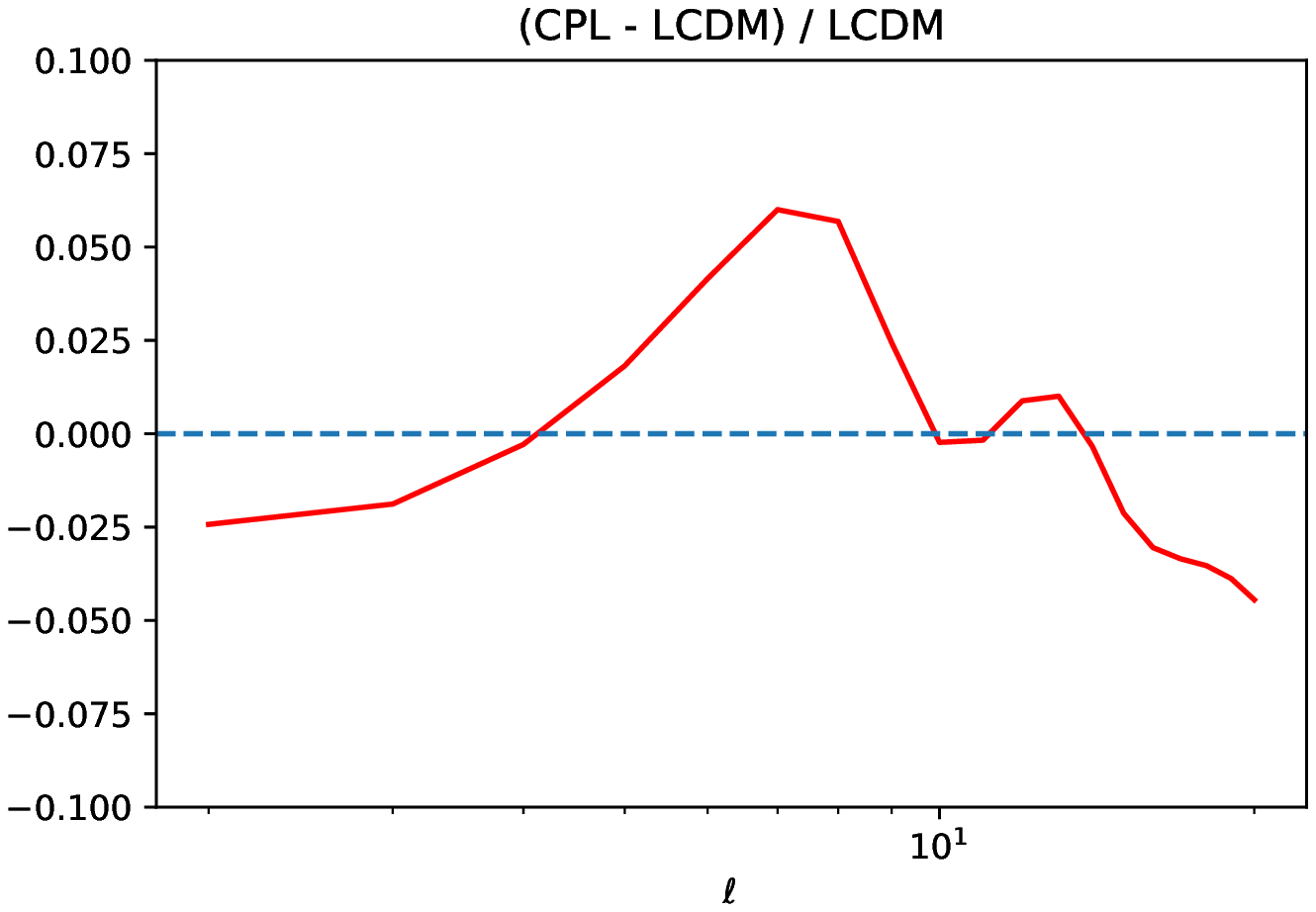}
\includegraphics[width=40mm]{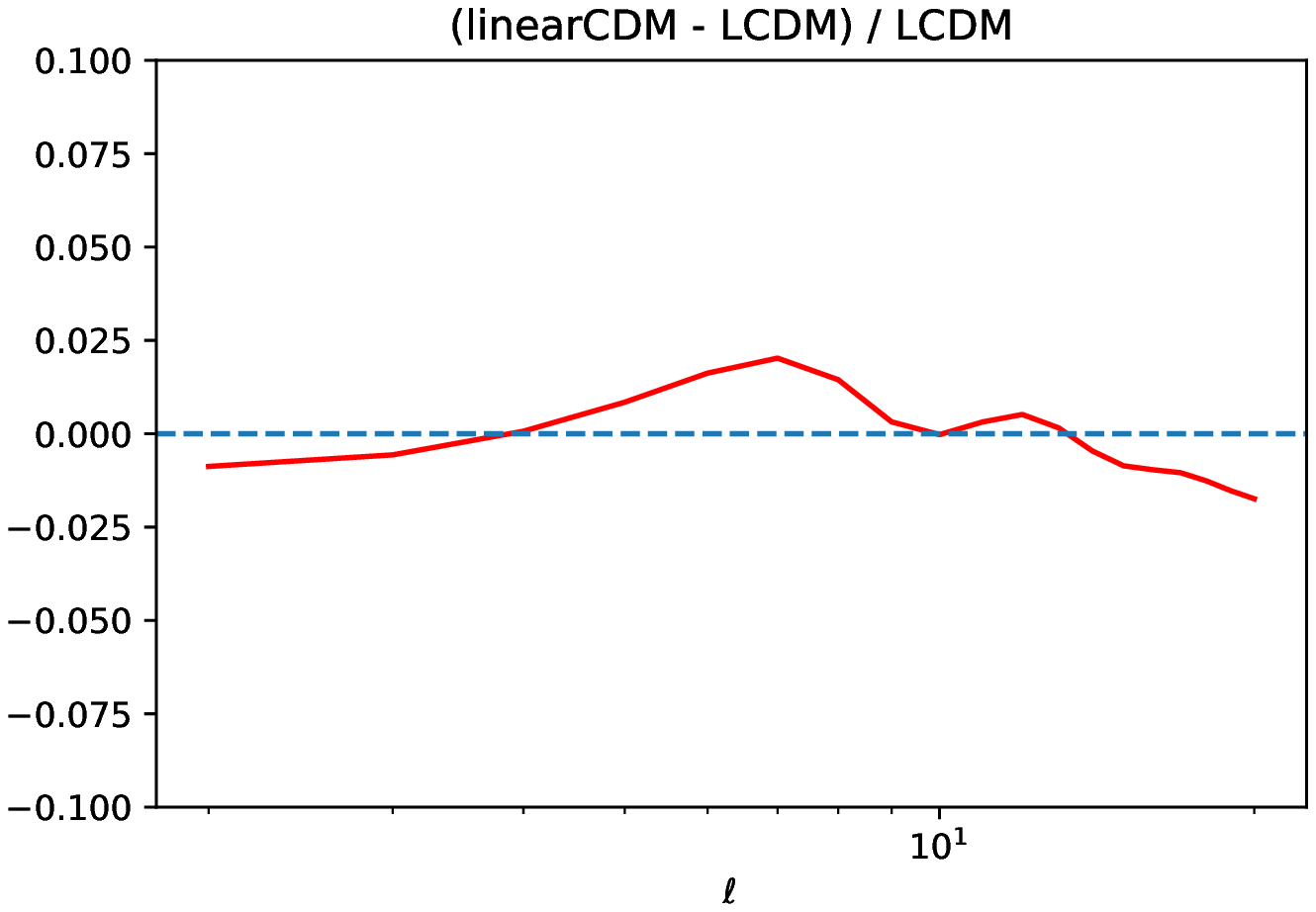}
\includegraphics[width=40mm]{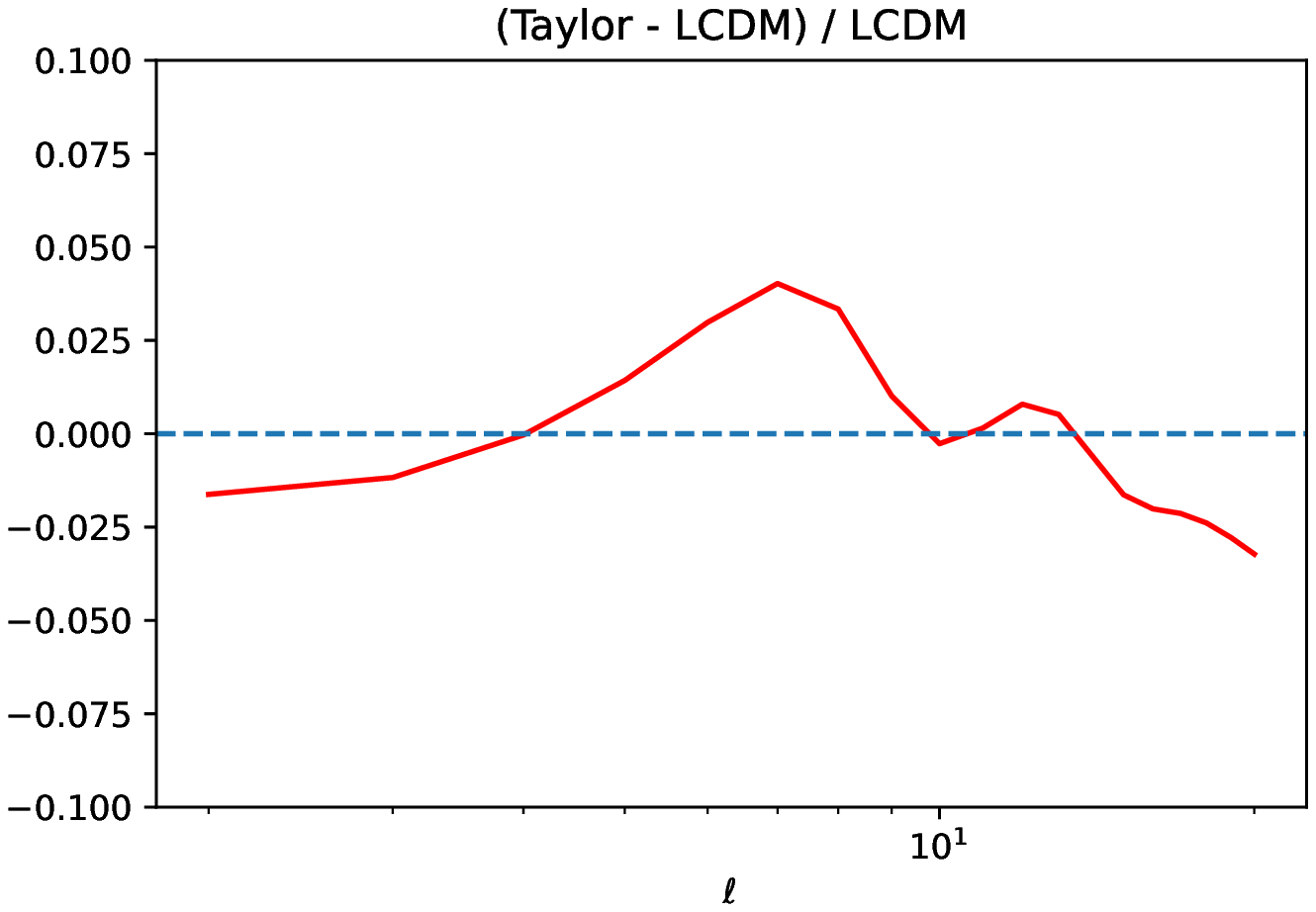}
\caption{
The same of Fig.~\ref{fig:Pantheon-diff},
 but in each figure the value of $\Omega_m$ in the $\Lambda$CDM model
 is artificially set the value of each corresponding model.
}
\label{fig:Pantheon-diff2}
\end{figure}

Since the difference of the EE polarization power spectrum of each model
 from that of the $\Lambda$CDM model
 is almost horizontal shift in $\ell$ direction,
 these results may be understood as the results of change
 in the angular diameter distance to reionization.
The angular diameter distance
 is proportional to an integration of the inverse of eqs.(\ref{E-CPL}) and (\ref{E-Taylor}),
 and the effect of each model is divided into two origins:
 the change of the value of $\Omega_m$ and
 the non-trivial redshift dependence of dark energy.
Fig.~\ref{fig:Pantheon-diff2} shows
 the differences of EE polarization power spectra
 between in each phenomenological model and $\Lambda$CDM model
 with the value of $\Omega_m$ in each corresponding model.
We see that
 the magnitude of the difference is reduced by about a factor of ten
 and the original difference is in fact dominated by the change of the value of $\Omega_m$.
The effect of the non-trivial redshift dependence of dark energy
 remains, even though it is small, which keeps the original shape,
 or $\ell$-dependence, of the difference.
Note, however, that
 the origin of the different value of $\Omega_m$ in each model from that of the $\Lambda$CDM model
 is the non-trivial redshift-dependence of dark energy
 in the fit with supernova redshift-magnitude relation. 

\begin{figure}[t]
\centering
\includegraphics[width=40mm]{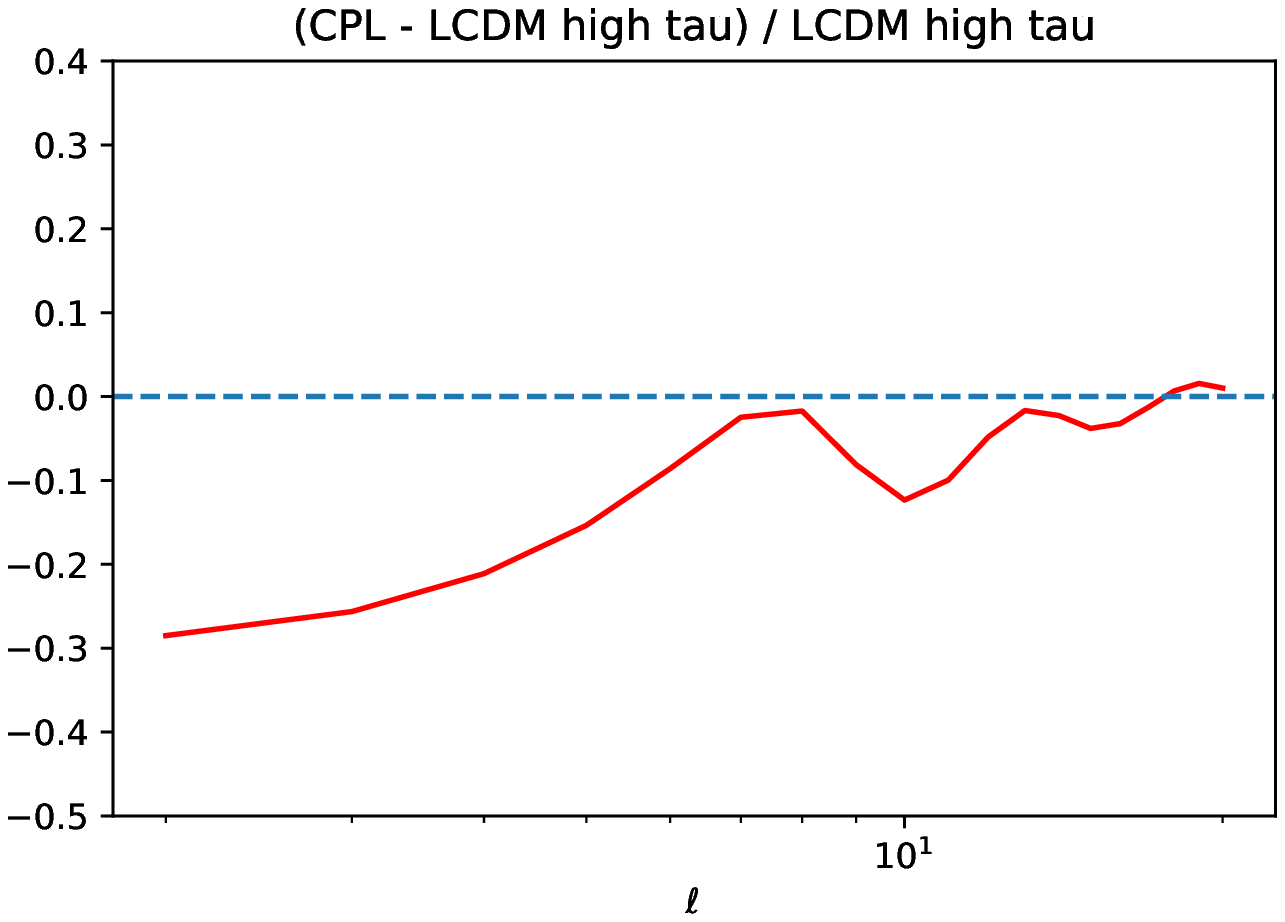}
\includegraphics[width=40mm]{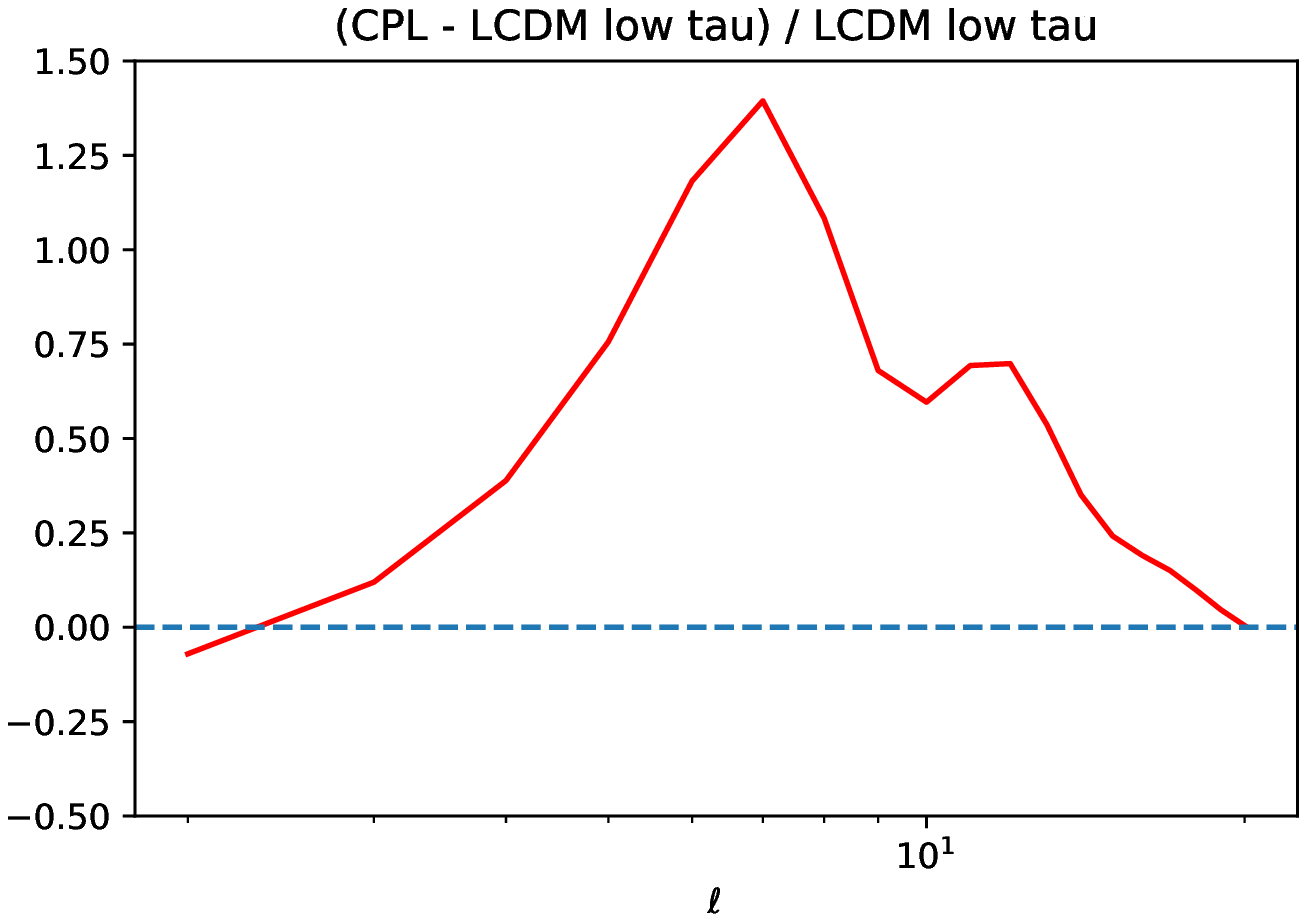}
\includegraphics[width=40mm]{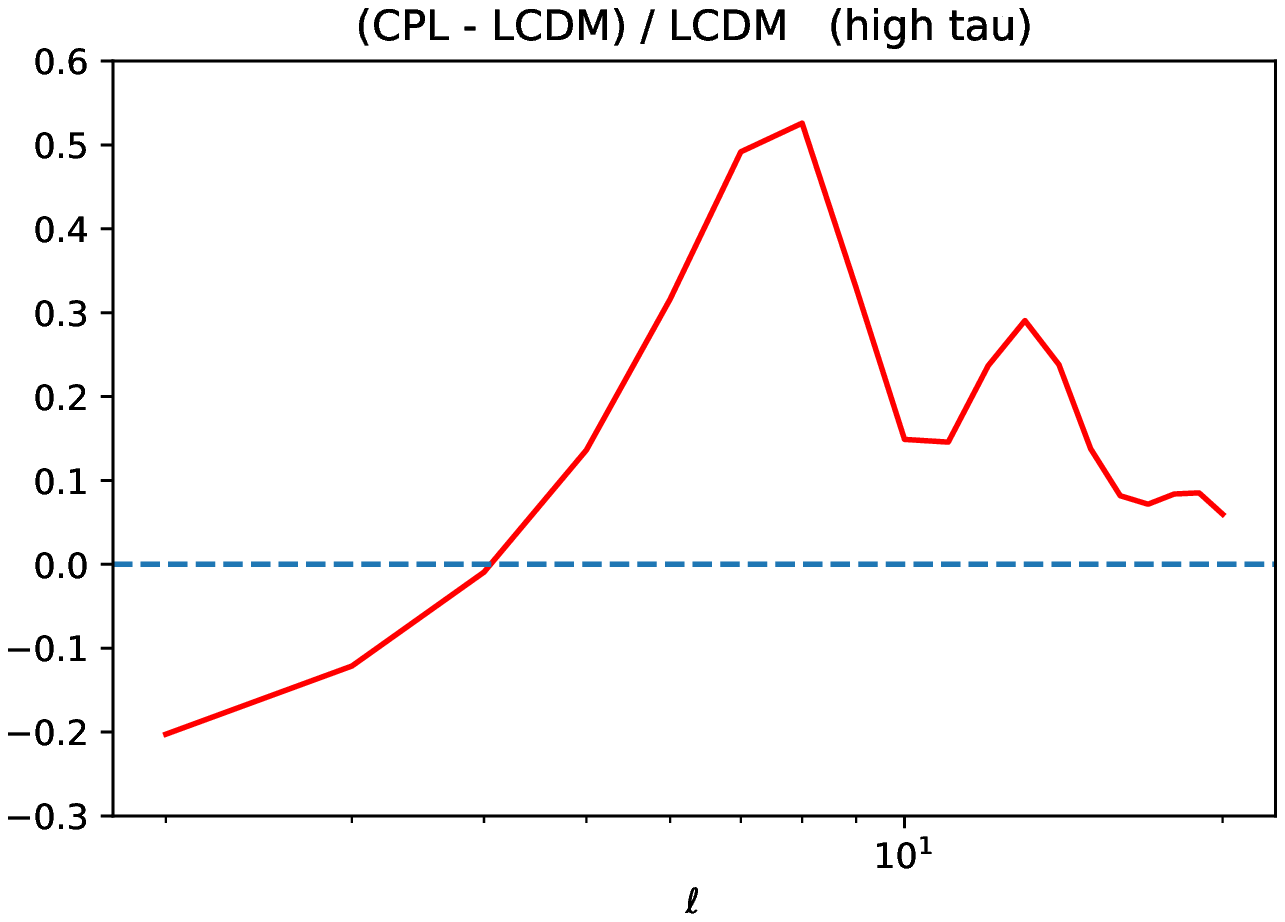}
\includegraphics[width=40mm]{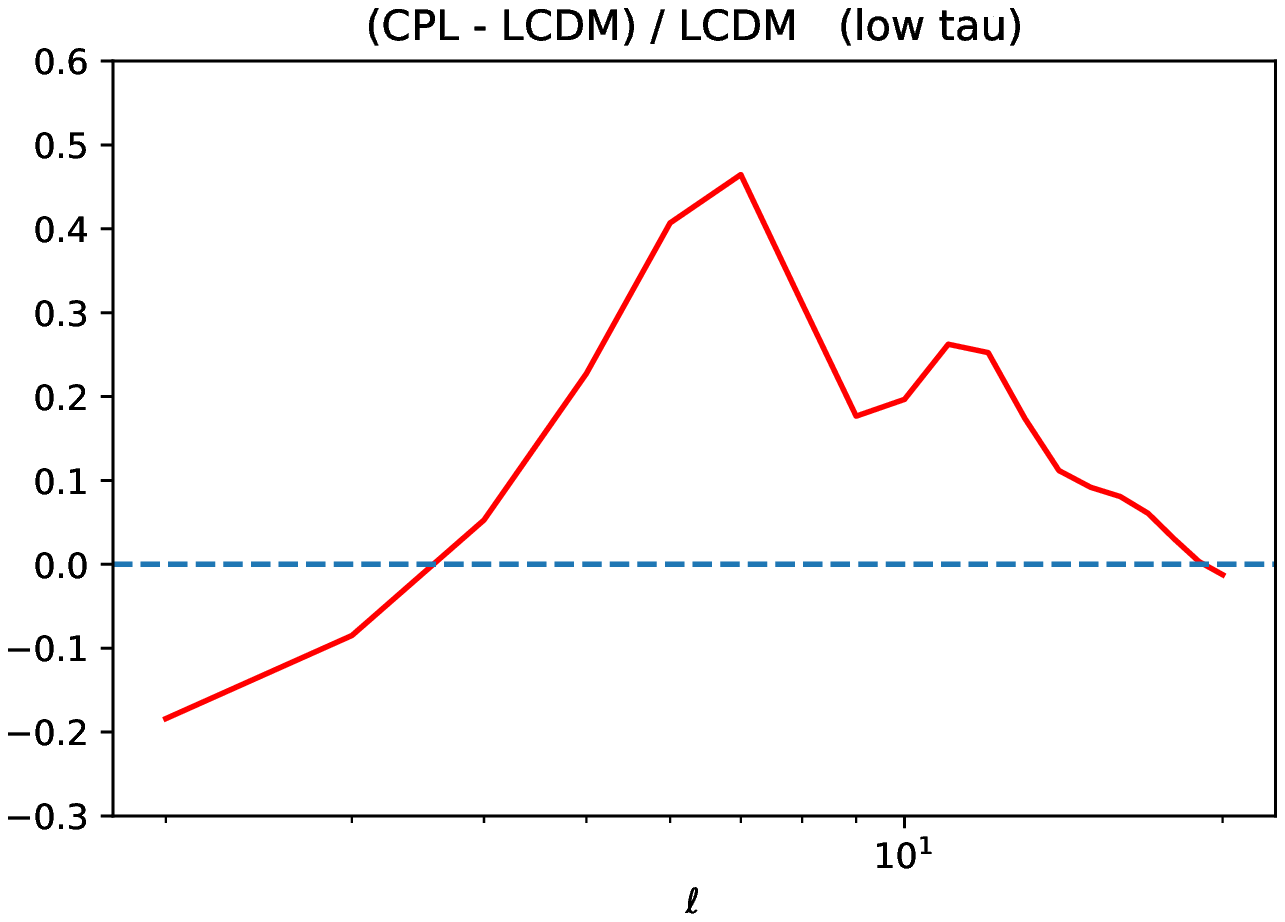}
\caption{
The differences of EE polarization power spectra
 between in CPL model and the $\Lambda$CDM model.
In the left figure the value of the optical depth is set as $\tau=0.054+0.007$ in the $\Lambda$CDM model
 and in the middle-left figure it is set as $\tau=0.054-0.007$ in the $\Lambda$CDM model
 keeping as $\tau=0.054$ in CPL model in both figures.
In the middle-right and right figures,
 the value of the optical depth is set as $\tau=0.054+0.007$ and $\tau=0.054-0.007$,
 respectively, in both CPL model and the $\Lambda$CDM model.
}
\label{fig:Pantheon-diff3}
\end{figure}

Here, we show the necessity and importance of the precise knowledge of the reionization process.
The typical present bound
 on the optical depth is $\tau = 0.054 \pm 0.007$ \cite{Aghanim:2018eyx}.
The two figures in Fig.~\ref{fig:Pantheon-diff3} from the left
 show the differences of EE polarization power spectra
 in case of $\tau=0.054+0.007$ and $\tau=0.054-0.007$ in the $\Lambda$CDM model, respectively,
 keeping $\tau=0.054$ in CPL model.
These figures show that
 the enhancement in the prediction of CPL model with $\tau=0.054$ is disappeared,
 if the $\Lambda$CDM model with $\tau=0.054+0.007$ is correct in Nature,
 and the opposite happens, if the $\Lambda$CDM model with $\tau=0.054-0.007$ is correct in Nature.
It is clear that the precise understanding about the reionization process is certainly necessary.
Fortunately,
 we can expect that the knowledge of reionization process will be extensively improved
 by observing 21-cm background by SKA \cite{SKA:2018ckk} and other probes, for example,
 before the cosmic variance limited measurements of low-$\ell$ EE polarization power spectrum
 by LiteBIRD.
Since the detailed quantitative evaluation of the uncertainty
 by ambiguous knowledge of reionization, like the work in \cite{Millea:2018bko},
 by forecasting the future ambiguity of the knowledge with various forthcoming observations
 is beyond the scope of this article, we leave it for future works. 
One naive possibility, which would be worth to be investigated in future, would be that
 the theoretical prediction of the difference of EE polarization power spectra
 between in some dark energy model and the $\Lambda$CDM model would not be affected so much
 by the detail of the reionization process
 (see the right two figures in Fig.~\ref{fig:Pantheon-diff3}).

We introduce the following two quantities
 to investigate the detectability of this enhancement.
One is the {\it variance} used in
 \cite{Monteserin:2007fv,Cruz:2010ud,Gruppuso:2013xba,Gruppuso:2017nap}
\begin{equation}
 V_{\rm model} \equiv
  \sum_{\ell=\ell_{\rm min}}^{\ell_{\rm max}} \left(\frac{2\ell+1}{4\pi}\right) C^{EE, {\rm model}}_\ell,
\label{variance}
\end{equation}
 which defines autocorrelation $C^{EE, {\rm model}}(\theta=0)$
 between $\ell_{\rm min}$ and $\ell_{\rm max}$.
The error of this quantity is defined as
\begin{equation}
 \delta V^2 \equiv
  \sum_{\ell=\ell_{\rm min}}^{\ell_{\rm max}}
   \left( \left(\frac{2\ell+1}{4\pi}\right) C^{EE, {\rm model}}_\ell \sqrt{\frac{2}{2\ell+1}} \right)^2,
\end{equation}
 where we assume that the error is cosmic-variance-limited.
We define the significance of enhancement as
\begin{equation}
 \Delta V_{\rm model} \equiv \frac{V_{\rm model}-V_{\Lambda{\rm CDM}}}{\delta V}.
\end{equation}
In this article we take $\ell_{\rm min}=4$ and $\ell_{\rm max}=20$.
Another quantity is the measure of the differences in $D^{EE}_\ell$ 
\begin{equation}
 \chi^2_{\rm model} \equiv
  \sum_{\ell=\ell_{\rm min}}^{\ell_{\rm max}}
  \frac{(D^{EE, \Lambda{\rm CDM}}_\ell - D^{EE, {\rm model}}_\ell)^2}{\sigma_\ell^2},
\label{chi2-model}
\end{equation}
 where
\begin{equation}
 \sigma_\ell \equiv D^{EE, \Lambda{\rm CDM}}_\ell \sqrt{\frac{2}{2\ell+1}}
\label{D-error}
\end{equation}
 is the error due to the cosmic variance only.
This second quantity
 is simply the measure of the difference in $\ell$ by $\ell$ in comparison with error,
 or a sort of {\it goodness-of-fit} of the results in $\Lambda$CDM model and phenomenological model.

\begin{table}[t]
\centering
\caption{
The quantities
 to measure the detectability of the enhancement in EE polarization power at low-$\ell$
 in case of the Pantheon binned data.
The degrees of freedom is defined as
 $\nu \equiv (\ell_{\rm max}-\ell_{\rm min}+1) - ({\rm number \ of \ model \ parameter)}$.
}
\label{table:V-and-chi2-binned-data}
\begin{tabular}{lccc}
\hline
 Model & $\Delta V_{\rm model}$ & $\chi^2_{\rm model}$ & $\chi^2_{\rm model}/\nu$ \\
\hline\hline
 $w$CDM       & 0.447 & 1.28 & 0.0800 \\
 CPL          & 1.07 & 8.45 & 0.563 \\
 linearCDM    & 0.446 & 1.29 & 0.0807 \\
 Taylor       & 0.763 & 4.03 & 0.269 \\
\hline
\end{tabular}
\end{table}
The quantities $\Delta V_{\rm model}$, $\chi^2_{\rm model}$ and $\chi^2_{\rm model}/\nu$
 ($\nu$ is the degrees of freedom) in case of the Pantheon binned data fits
 are summarized in Table~\ref{table:V-and-chi2-binned-data}.
We see that the enhancement is small in comparison with error in each model.
The differences of {\it variances} are less than $1\sigma$,
 and {\it fits} between $D^{EE, \Lambda{\rm CDM}}_\ell$ and $D^{EE, {\rm model}}_\ell$
 are rather good.
The effect in CPL model appears larger than those in linearCDM and Taylor expansion models.
This would indicate that
 the extrapolation of the CPL model to larger redshift region could be problematic,
 because the expansion in $(1-a)=z/(1+z)$ may not be valid at large $z \sim 10$,
 though $a = 1/(1+z)$ remains small in Taylor expansion model.
It will be difficult
 to measure this enhancement by the non-trivial time dependence of dark energy
 even in future precise measurements of CMB EE polarization power spectra
 by LiteBIRD \cite{Hazumi:2021yqq}, for example.
Furthermore,
 since this phenomenon rather strongly depends on the knowledge of reionization process
 (the value of optical depth $\tau$, as we have seen, and reionization process
 (see \cite{Kitazawa:2020qdx} for quasi-quantitative arguments)),
 it will be required a global analysis relying on the consistency
 between many observables, including which will be available in future,
 not simply looking the low-$\ell$ EE polarization power spectrum only.

\section{Conclusions}
\label{sec:conclusions}

It has been shown that the redshift-magnitude relation of type Ia supernova
 still allows various non-trivial time-dependences of the equation of state of dark energy.
The $\Lambda$CDM model is not necessary the best model from the statistical point of view.
Though the analysis in this article is rather simple one,
 some recent rigorous analysis give similar results:
 the constraints on the parameters is not strong enough
 in \cite{Alam:2020sor} and also \cite{Abbott:2021bzy} more recently, for example.

It would be worth to find other physical quantities
 which can give constraints to the time dependence of the equation of state of dark energy.
We have investigated the possibility that
 the CMB polarization power spectrum at low-$\ell$ (large scales) can be such a quantity.
This is because that the non-trivial time dependence of dark energy
 changes the way of expansion of the universe in the period of reionization,
 mainly through the change of the value of the matter density parameter $\Omega_m$
 due to the fit with the redshift-magnitude relation of type Ia supernova,
 and the low-$\ell$ polarization is produced by the Thomson scattering
 with free electrons which are produced in the reionization process.

We have found that
 the CMB EE polarization power spectrum is enhanced around $4 < \ell <20$
 in comparison with the prediction of $\Lambda$CDM model in rather model independent way.
However,
 the amount of enhancement is rather smaller
 in comparison with the ambiguity or error due to cosmic variance.
Therefore,
 it will be difficult to constrain non-trivial time dependence of dark energy
 by simply looking the low-$\ell$ EE polarization power spectrum only.
In case that the tensor-to-scalar ratio would be large
 and BB polarization power spectrum at low-$\ell$ would be observed in future,
 it would be an important observable to constrain the non-trivial time dependence of dark energy.
The future efforts
 to understand the reionization process more precisely (see \cite{Watts:2019uvq} in CMB, for example)
 are certainly very important.
The future information on the cosmology at high redshift ($6 < z < 10$)
 by 21-cm and other observations will be very important also.
Especially,
 the expansion history of the universe at high redshift
 is very important for clarifying the nature of dark energy
 (see \cite{Sailer:2021yzm} for future prospects, for example).

The same analysis using different supernova catalog,
 the Foundation Supernova Survey \cite{Foley:2017zdq,Jones:2018vts}, for example,
 would be very interesting to clarify whether
 the results of this article include some systematic error related with supernova data.
A characteristic feature of the Foundation Supernova Survey is that
 they focus on the data of redshift region $z<0.1$ where the universe is dark energy dominant.

\section*{Acknowledgments}

This work was supported in part by JSPS KAKENHI Grant Number 19K03851.

\end{document}